# Astro2020 Science White Papers

# Increasing the Discovery Space in Astrophysics
# A Collation of Six Submitted White Papers

**Included Thematic Areas:**   Formation and Evolution of Compact Objects
Multi-Messenger Astronomy and Astrophysics
Stars and Stellar Evolution
Planetary Systems
Star and Planet Formation
Resolved Stellar Populations
Galaxy Evolution
Cosmology and Fundamental Physics

**Principal Author:**
Name: G. Fabbiano
Institution: Center for Astrophysics | Harvard & Smithsonian (CfA)
Email: gfabbiano@cfa.harvard.edu

**Co-authors:**
M. Elvis (CfA), A. Accomazzi (CfA), A. Avelino (CfA, Harvard), G. B. Berriman (Caltech/IPAC-NExScI), N. Brickhouse (CfA), S. Bose (CfA), D. Carrera (Penn State Univ.), I. Chilingarian (CfA), F. Civano (CfA), B. Czerny (Center for Theoretical Physics, Warsaw; CfA collaborator), R. D'Abrusco (CfA), B. Diemer (Cfa), J. Drake (CfA), R. Emami-Meibody (CfA), J. R. Farah (CfA), G. G. Fazio (CfA), E. Feigelson (Penn State Univ.), F. Fornasini (CfA), Jay Gallagher (Univ. Wisconsin), J. Grindlay (CfA), L. Hernquist (CfA), D. J. James (CfA), M. Karovska (CfA), V. Kashyap (CfA), D.-W. Kim (CfA), G. M. Lacy (NRAO), J. Lazio (JPL, Caltech), E. Lusso (Univ. of Florence, Italy; CfA collaborator), W. P. Maksym (CfA), R. Martinez Galarza (CfA), J. Mazzarella (Caltech/IPAC), M. Ntampaka (CfA), G. Risaliti (Univ. of Florence, Italy; CfA collaborator),D. Sanders (Institute for Astronomy, Hawaii), N. Scoville (Caltech), I. Shapiro (CfA), A. Siemiginowska (CfA), A. Smith (STScI), H. Smith (CfA), I. Stephens (CfA), A. Szentgyorgyi (CfA), S. Tacchella (CfA), A. Thakar (Johns Hopkins Univ.), V. Tolls (CfA), S. Vrtilek (CfA), B. Wilkes (CfA), D. Wilner (CfA), S. P. Willner (CfA), S. J. Wolk (CfA) Zhao, J.-H. (CfA)

## Abstract

We write in response to the call from the 2020 Decadal Survey to submit white papers illustrating the most pressing scientific questions in astrophysics for the coming decade. We propose exploration as the central question for the Decadal Committee's discussions. The history of astronomy shows that paradigm-changing discoveries were not driven by well-formulated scientific questions, based on the knowledge of the time. They were instead the result of the increase in discovery space fostered by new telescopes and instruments. An additional tool for increasing the discovery space is provided by the analysis and mining of the increasingly larger amount of archival data available to astronomers. Revolutionary observing facilities, and the state-of-the-art astronomy archives needed to support these facilities, will open up the universe to new discovery.

This white paper includes science examples of the power of the discovery approach, encompassing all the areas of astrophysics covered by the 2020 Decadal Survey.

## Table of Contents

# 1. The Exploration Question

There has been a long-standing tension in our discipline between the 'exploration' approach and the more physics-based 'question-driven' approach.

### 1.1 - *The **question-driven** approach*

This approach seeks to formulate the most important open questions in our discipline. It is based on our *present knowledge* of the field (both theoretical and observational) and is formulated usually as a way to constrain and/or advance a currently proposed cosmological or astrophysical scenario. 'Question/hypothesis-driven' has been the preferred approach in the last few decades and is used to justify both observing proposals and proposals for new instruments and telescopes. Most talks at conferences and papers are framed based on this question and answer approach. This is how we teach our students to approach research. This is the approach formulated in the Decadal Survey call for white papers. This approach addresses the '*known unknowns*': for example, the way to best constrain the cosmological parameters of our universe, and lately the search for dark matter and dark energy, and the definitive discovery of gravitational waves.

The question-driven approach continues to be fruitful, and it gives us a certain sense of control in our progress, but -by its own nature- is also a limited and limiting epistemology. For example, it can bias our knowledge. As expounded in a recent article with reference to extra solar planets "the key is to make sure that science policy permits discovery for the sake of discovery and not for finding Earth-like planets, which we have prejudiced to be of greatest interest (D. J. Stevenson, CalTech, Physics Today, Nov. 2018)". The same opinion can be easily shaped to apply to other fields of astrophysics.

The question-driven approach does not address the *'unknown unknowns'* that by their nature cannot be addressed as well-defined 'important questions'.

### 1.2 - *The **exploration** approach*

This approach, i.e. gaining the capability to find new questions, rather than solving known ones, is the only way we can address the *unknown unknowns*. Harwit (1984) calls this '*discovery space*'. The notion that most of science is undiscovered and that 'out of the book' thinking may be needed for real progress is making fast inroads (e.g., see the book 'Ignorance: How it Drives Science' by S. Firestein, 2012). How to best foster the discovery of *unknown unknowns* is particularly poignant for astronomy, which throughout its history has been first and foremost exploratory.

The real big paradigm shifts in astronomy and astrophysics have occurred when new approaches have significantly opened up the discovery space, revealing unforeseen views of the universe. These approaches may have been framed as a way to address important questions of the time, but the real advances were from serendipitous discoveries. The discovery space may have been increased by means of new telescopes and instruments (both hardware and software), and also by *unanticipated data repurposing*.

Famous examples of discoveries stemming from exploration include:

- The Galilean Moons of Jupiter, the metal composition of the Sun and stars, the HR diagram, the expansion of the Universe, large scale structure, hot Jupiters (driven by improvements in optical telescopes and spectrographs);

- Quasars, radio galaxies, the microwave background, pulsars, superluminal motion, fast radio bursts (following the invention of radio telescopes, VLBI, and search in the archives in the case of bursts);
- Black holes and their mass range, dark matter, dark energy, super-starburst galaxies (from the availability of new space-based observing windows, X-ray, IR, and high resolution optical imaging with HST, and availability of multi-wavelength archives).

*These foundational discoveries for the present understanding of the Universe and its evolution were not in any way anticipated.* Most of them were fostered by the use of increasingly larger telescopes and more sensitive instruments, able to explore different parts of the electromagnetic spectrum. Others were surprising results of the data analysis.

Given the increasing availability of large and survey data sets in our open archives, a new hybrid approach, **question-driven exploration**, has emerged, where astronomers have mined these data and researched the literature guided by relatively vague questions, finding answers, new questions, and surprises. A similar approach is making inroads in biology (Elliott et al 2016).

In this white paper we discuss the '*exploration question',* providing examples relevant for the fields covered by the 2020 Decadal Survey. Given the nature of exploration, it is not possible to give definite questions that need to be addressed in the near future. Rather, we provide a few examples of (1) serendipitous unexpected discoveries (*unknown unknowns*) and their potential for changing established paradigms; and (2) new research avenues posed by asking very general questions (*known unknowns*). We do not mean to provide an exhaustive survey of such discoveries, but only to illustrate our case with a few representative studies (Sections 2 - 7). In Section 8, we address our recommendations for increasing the discovery space, and in Section 9 we give high-level conclusions.

## 2. Exploration in Compact Objects

Although there were a few theoretical predictions beforehand, compact objects are excellent exemplars of discoveries made through exploration. In the space of a few years, *neutron stars* were found both as pulsars, seen as "scruff" in radio survey data (Hewish, Bell et al. 1968), and as bright periodic X-ray sources; *black holes* were found as rapidly variable X-ray sources and, later, dynamically with spectroscopy of the secondaries of faded X-ray transients (McClintock and Remillard 1986). *Gamma-ray bursts* discovered at the same time (Klebesadel et al., 1973) were mysterious for decades but also implied compact objects. *Fast radio bursts* are a more recent discovery and may be connected with compact objects.

In addition to being *totally unexpected discoveries*, compact objects (including stellar-mass black holes and neutron stars), required observations at wavelengths other than the wavelength of their discovery (typically X-ray, radio or gamma-ray) before their true nature could be determined. An explanation for radio pulsars was developed within months of their discovery (Gold 1968), for X-ray binaries within years of discovery (Pringle & Rees 1972), and for gamma-ray bursts we are only now, more than half a century later, beginning to understand their nature:

- Radio pulsars provided us with the first (indirect) evidence for the existence of *gravitational waves*. Pulsar glitches led to the recognition of solid crusts on *neutron stars*, and neutron stars offer ways to test the nature of the strong nuclear force via their equation of state.

- X-ray binaries use the most efficient means known of extracting energy from matter. Their messy phenomenology revealed its hidden order when color-color and color-intensity diagrams from archival data showed clear paths between different accretion states (Done & Gierlinski 2003), reflecting the balance of disk and jet emission as accretion rates change.
- The known mass range of stellar *black holes* was greatly extended by LIGO *gravitational wave* detections (Abbott et al 2017a). Joint observations of gravitational wave signal GW170817 by the LIGO-Virgo detector network and transient electromagnetic counterparts from multiple telescopes resulted in a triumph for multi-messenger astronomy: discovery of an inspiraling binary neutron star (Abbott et al. 2017b).
- Bright X-ray bursts (discovery paper, Grindlay, Gursky et al 1976) are energetic phenomena typical of accreting neutron stars in binaries, associated with thermonuclear flashes on the NS surface (see e.g., review Lewin et al 1995).
- Gamma-ray bursts, which are the most energetic phenomena known to humanity, have been connected with neutron star mergers, resulting in the first *multi-messenger* observations of *gravitational wave* events (e.g., Haggard, et al., 2017). Multi-wavelength light curves can provide a physical view of the phenomenon (Kasliwal et al. 2017).
- Fast Radio Bursts are transient phenomena lasting only milliseconds, are suspected of being compact objects, and many are being discovered by an instrument designed to look for redshifted 21cm hydrogen emission (Chime/FRB Collaboration 2019). There are dozens of ideas about what they *might* be, but no way to discern between them. FRBs may provide new cosmological tests (Jaroszynski, 2019).

The archival data available at different wavelengths provide a resource for exploration, both for identifying new phenomena that may be related to compact objects, finding precursor objects, and investigating systematics in newly identified classes. For example:

- A blind search for pulsations of 13 years of XMM observations containing 50 billion photons led to the discovery of the first *extragalactic X-ray pulsar* in a globular cluster and the first pulsar in the Andromeda galaxy, which is also the second slowest X-ray pulsar known (p=1.2sec, Zolotukhin et al., 2017). This long period challenges theories of binary star formation and evolution in clusters, as it must have started accreting recently, about 1 Myr ago, yet is hosted in a 12 Gyr-old globular cluster.
- Archival searches for *fast radio bursts* are ongoing and beginning to produce results (e.g., Zhang et al., 2019).
- Spectral and timing properties of populations of X-ray sources discovered with *Chandra* in nearby galaxies have been used to compare them with the known behavior of compact Galactic binaries. Color-color-intensity diagrams were used to separate classes of X-ray binary (black holes and neutron stars) and to separate jet-producing versus non-jet-producing XRBs. This is directly akin to color-magnitude, Hertzprung-Russell diagrams for stars, which was also unanticipated (Vrtilek & Boroson 2013).
- Comparison with *Hubble* imaging and photometry have led to extraction of globular cluster binaries; association of binaries with different stellar population age and metallicity; and even to the realization that both globular clusters and X-ray binaries may trace merging accretion events in giant elliptical galaxies. These comparisons have led to constraints on the nature of these sources and their evolution (e.g., see reviews: Fabbiano 2006; 2019).

The identification of fast timing events at different wavelengths will also require *archival databases* for studying the baseline state and past history of the associated object. As a case in point, researchers racing to find an optical counterpart to binary neutron star merger GW170817 made use of NED to point the Swope telescope and discover an afterglow in lenticular galaxy NGC 4993 only 11 hours after the GW trigger (Coulter et al. 2017); this led to development of a new service[1] to facilitate follow-up observations of GW events. New computational techniques powerful enough to process TB/day rates of high-dimensional data with low latency will be important to *multi-messenger* astronomy in the LSST plus LIGO+, VIRGO, KAGRA, INDIGO era.

## 3. Exploration in Stars and Stellar Evolution

Given the nature of exploration, it is not possible to give definite questions that need to be addressed in the near future. Rather, we provide a few examples of (1) serendipitous unexpected discoveries (*unknown unknowns*) and their potential for changing established paradigms; and (2) new research avenues posed by asking very general questions (*known unknowns*). We do not mean to provide an exhaustive survey of such discoveries, but only to illustrate our case with a few representative studies.

### 3.1 *Unanticipated discoveries from Hubble and other space observatories*

Past the first discovery period that defined stellar physical and chemical properties, and spurred the first theoretical work on the stellar engine and stellar evolution, stellar astronomy / astrophysics is an area where discovery guided by theoretical models has been predominant and works well. Despite this, the advent of observations with *Hubble*, with its high angular resolution, has provided unexpected discoveries, for example (1) the complex stellar populations of globular clusters (Gratton et al. 2012; Piotto et al. 2015), and (2) the major role of binary star evolution in mass ejection, e.g., in stellar systems ranging from the production of planetary nebulae to binary star mergers (Jones & Boffin 2017; Smith et al. 2016, Tylenda & Soaker 2006).

New generations of *multi-wavelength data* have led to new insights into stellar physics. The enhanced capabilities provided by space observatories fostered multi-wavelength stellar research leading to unanticipated results, such as:

- The detection of black hole transients, and routinely finding dusty disks accreting matter onto protostars (with an accretion shock detected with *Chandra* in TW Hya), and producing jets (Burrows et al. 1996, McCaughrean & O'Dell 1996, Kastner et al 2002; Prisinzano et al. 2008).

- The detection of stellar oscillations, especially in evolved red giant stars by *CoRoT* and *Kepler*, which opened a powerful avenue to test stellar models (Chaplin & Miglio 2013),

- The detection of γ-rays from novae in outburst, which revealed their ability to act as shock-powered high-energy particle accelerators (Ackermann et al. 2014).

- The discovery of coronal activity in a nearby brown dwarf with *Chandra*, showing that these ultra-cool objects undergo energetic reconnection flares similar to fusion-powered stars like the Sun (Rutledge et al. 2000).

---

[1] https://ned.ipac.caltech.edu/gwf/

The ability to connect a variety of multi-wavelength observations to sophisticated models is a key factor behind these advances. Progress thus depends on access to data by researchers with a wide variety of backgrounds.

**3.2** *New understanding from archival studies*

New research avenues are opened by the presence of the growing amount of stellar data available in astronomy archives, and by the availability of powerful software tools to exploit these data. In these types of projects exploration comes from the capture of high quality data for large samples of objects. Hundreds of thousands of high quality stellar spectra have been collected by different observatories over the last three decades and put into data archives. Tools to determine stellar atmospheric parameters (effective temperature, surface gravity, elemental abundances, rotation) evolved over time and became mature only a few years ago thanks in particular to the development in the field of studies of exoplanets. New methods of data exploration are also coming to the fore, especially based on classification and clustering based on machine learning. Examples of this approach include:

- AMBRE, a project that re-processed some 52,000 high-resolution stellar spectra collected with 4 different instruments at the European Southern Observatory (Worley et al. 2012; de Laverny et al. 2013). By combining precise measurements of stellar parameters and radial velocities obtained from spectra with *Gaia* information on distances and proper motions, archival studies in the area of Galactic astronomy have an enormous potential in the next decade. Analyzing the archival stellar spectra in a uniform way and determining stellar properties will help to pin down some of the unsolved questions of stellar evolution and synthesis of heavy elements in stellar interiors.

- The discovery of a diagonal rotation period gradient across the main sequence in stars from *Gaia* Data Release 2, matched with rotation periods from *Kepler*, which may be due to metallicity effects (Davenport & Covey 2018; Davenport 2019).

A variety of stellar spectroscopic surveys are in progress or planned for the next decade, with a variety of goals. These surveys will provide huge accessible archival databases. They include:

- APOGEE, a program in the Sloan Digital Survey IV designed to study stellar abundances patterns in the Milky Way.

- ULLYSES, the recently announced project from STScI that seeks to produce a full library of ultraviolet spectra of massive stars in the SMC, to provide a knowledge base for modeling stars at low metallicity. This project will gain support from a number of PI-programs of optical spectroscopy of massive stars in the SMC, such as those that were enabled by the capabilities at ESO for moderate-to-high spectral resolution multi-object observations over substantial fields of view.

These studies will further benefit from connections to future multi-messenger surveys such as the photometric studies by WFIRST and SPHEREx, as well as new generations of gravity wave and neutrino observatories. The results will impact extragalactic astronomy because new generations of stellar population models can be created utilizing improved stellar astrophysics and used to quantitatively explore the evolution of galaxies.

The time domain also is an area where enhanced capabilities led to new discoveries and where more progress will come, e.g., with the advent of LSST. Examples include the OGLE gravitational lensing surveys that led to new understanding of the structure of the Magellanic Clouds. Many optical transient surveys have enriched greatly our understanding of supernovae and stellar variability. Key features of these programs include systematic approaches to the capture, analysis, and delivery of results to the community.

## 4. Exploration in Planetary Systems and Planet Formation

Given the nature of exploration, it is not possible to give definite questions that need to be addressed in the near future. Rather, we provide a few examples of (1) serendipitous unexpected discoveries (*unknown unknowns*) and their potential for changing established paradigms; and (2) new research avenues posed by asking very general questions (*known unknowns*). We do not mean to provide an exhaustive survey of such discoveries, but only to illustrate our case with a few representative studies.

**4.1** *Pulsar planets*

Aleksander Wolszczan (2000) discovered the first exoplanet and the first multiple exoplanet system (Konacki & Wolszczan 2003) in a place where nobody would ever have expected to find one: Orbiting a pulsar. This was done with Arecibo, a powerful instrument that was clearly not designed for the purpose of finding pulsar planets.

**4.2** *Hot Jupiters*

51 Peg b, the first exoplanet found orbiting a main sequence star, is a hot Jupiter. That is a type of planet that no theorist would have imagined could form. A week after its discovery, it was confirmed by looking at archival data from the Lick Observatory. This is a bit of a mixed discovery --- partly serendipitous, partly question-driven ---. Michael Mayor and Didier Queloz (1995), and other people were in fact trying to find exoplanets. But the thing they found was not at all what they expected.

**4.3** *Compact short period planet systems*

The *Kepler* mission set out to find the frequency of Earth-like planets. Although it sort of failed in that mission because the momentum wheels broke, *Kepler* was an incredibly successful mission because it found a whole new population of planets that we did not know even existed. *The Galaxy is littered with compact planetary systems with orbital periods well inside the orbit of Mercury.* This again, is a mix of "question driven" and "exploration driven" discovery.

*Kepler* was designed to answer one question, but its greatest contribution to the field was in answering a question that no one had asked.

**4.4** *Super-Earths and mini-Neptunes*

The open search for planetary transits by the Kepler mission led to the discovery that the most common class of planet is one that does not exist in our solar system – one that has a size larger than Earth, but smaller than Neptune (2-10 Earth Radii) and typically periods longer than three days. The nature of these novel planets is ambiguous as they lie near the boundary between rocky planets and gas giants. These planets may have their atmospheres stripped by high-energy particles from the host star (c.f. Fulton & Petigura 2018).

## *4.5* *Atmospheric Haze*

Heated and extended atmospheres have been detected around several Hot Jupiters and hot Neptunes. Photometry has suggested the existence of inhomogeneous Silicate-based cloud layers in several hot Neptune systems. Transmission spectroscopy has detected sodium and water vapor in the atmospheres of hot Jupiters, as well as <u>*unpredicted hazes*</u>, made of solid particles in an atmosphere that were formed by photochemistry.

## *4.6* *HL Tau and its protoplanetary disk*

HL Tau is a very young ($10^5$ years) T Tauri star surrounded by an equally young protoplanetary disk. Figure 1 shows an ALMA image of HL Tau. These observations were science verification taken by ALMA to test the long interferometric baselines. Nobody expected to see such substructure in a disk that is so young.

The image of HL Tau shows this very distinct ring structure and nobody's really sure of what it means. Recall that ALMA shows the location of the dust component of the disk. So the *dust* is in rings. Why? Could the black regions be gaps caused by planets? That is hard to believe given how young the system is and how many gaps there are. Maybe the dust is being accumulated inside pressure bumps in the gas that are caused by… something? Or perhaps there are no pressure bumps… perhaps the gaps are the signature of dust being converted into planetesimals… maybe. In any case, we still do not know what this image means, but we do know that HL Tau is not unique. ALMA has since found similar structures in many other protoplanetary disks (e.g., Andrews et al (2018); van der Marel et al. 2019), and there is now a great deal of effort in trying to understand what this all means.

HL Tau is again partly an example of question-driven and exploration-driven discovery. ALMA was intended to study protoplanetary disks. But nobody had expected such young protoplanetary disks to have this many rings. That discovery came as a by-product of the new capabilities provided by ALMA.

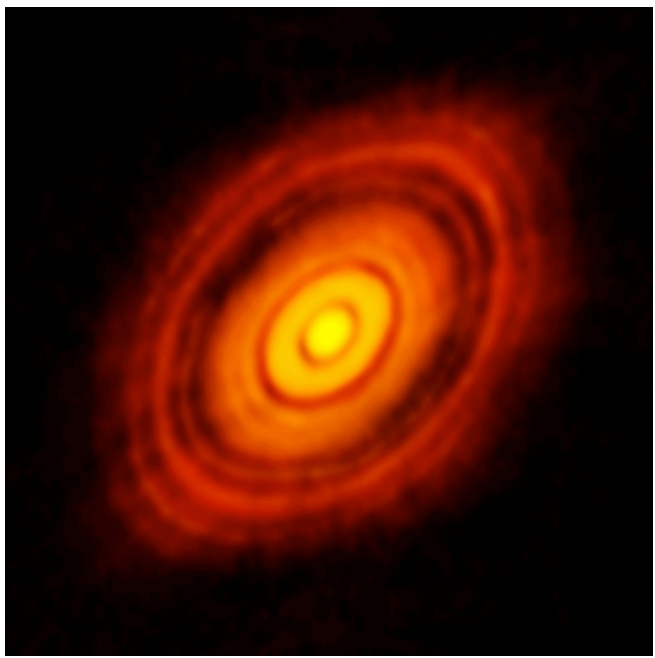

Figure 1. - ALMA (2015) image of the HL Tau disk

**4.7** *Protoplanetary disk ionization*

For decades, astrophysical models of the Solar Nebula and (later) protoplanetary disks treated them as cold structures of gas and dust with no external effects other than gravitational attraction and blackbody heating from their host star. But in 1991, Balbus & Hawley (ApJ, 3000 citations) realized that only $10^{-12}$ fractional ionization of cold molecular material is sufficient to initiate the *magneto-rotational instability*, rapidly building a magnetic dynamo and induce MHD turbulence. This solved some problems (providing a source of viscosity needed for accretion, and mitigating Type I inward migration of Jovian protoplanets) but raised other problems (inhibiting dust settling needed to initiate planetesimal growth). Hundreds of theoretical studies relating to planet formation in turbulent disks emerged, but also unexpected interactions with observations.

*Chandra,* NASA's flagship X-ray observatory, was studying variable X-ray emission in thousands of pre-main sequence stars. These empirical studies showed that hard X-rays from protostellar magnetic reconnection flares can penetrate deep into the disks, and probably dominate the ionization of the planet forming region. Thus, a telescope designed to elucidate superheated disks around black holes is surprisingly addressing the astrophysics of cold circumstellar disks giving rise to planetary systems.

# 5. Exploration in Resolved Stellar Populations

Given the nature of exploration, it is not possible to give definite questions that need to be addressed in the near future. Rather, we provide a few examples of (1) serendipitous unexpected discoveries (*unknown unknowns*) and their potential for changing established paradigms; and (2) new research avenues posed by asking very general questions (*known unknowns*). We do not mean to provide an exhaustive survey of such discoveries, but only to illustrate our case with a few representative studies.

**5.1** *The Panchromatic Hubble Andromeda Treasury (PHAT) survey*

PHAT (Dalcanton et al. 2012) consists of an imaging survey of ~1/3 of M31's star-forming disk in six filters, UV to NIR, to study resolved stellar populations in a controlled environment, at a given distance and metallicity while also avoiding the difficulties of observing stars through the inclined disk of our own Milky Way. PHAT resolved M31 into millions of individual stars, providing excellent constrains on stellar temperatures, bolometric luminosities, and extinction. *Unanticipated discoveries from the subsequent exploration of the dataset, and from combining it with other multi-wavelength surveys*, include:

- The discovery of weak CN stars, a previously unknown type of carbon star that appears to be associated with the He-burning phase of relatively massive stars (Masegian et al. 2019).
- The apparent universality of the high-mass (M>1M_sun) Initial Mass Function (IMF) across a broad range of cluster masses, ages, and sizes (Weizs et al. 2013),
- The strong UV bump in the extinction curve in the central region of M31, which indicates that dust destruction by supernova explosions are common in bulges of spirals (Dong et al. 2014).

**5.2** *Characterization of the structure of the Milky Way halo via data exploration.*

Unexpected discoveries about the morphology of the Galactic halo stem from the exploration and mining of several large data sets:

- Using astrometric data from *Gaia* (Lindegren et al. 2018), Prince-Whelan et al. (2018b) found a young, low-mass metal-poor stellar association while searching for groups of co-moving blue stars in the far end of the Galactic halo. Its age and location suggest that this association was formed when the leading arm of the Magellanic Cloud gas stream last encountered the M.W. disk.

- With a joint *Gaia* and Pan-STARRs study, Price-Whelan et al. (2018a) discovered an extension of GD-1, the longest known cold stream in the Galactic halo. They derived the position of the progenitor tidally disrupted globular cluster and detected over-densities in the stream that might be related to perturbation by the long-predicted dark-matter subhalos.

- Using Deimos spectra (Keck-II) and the Palomar Transient Facility Database, Cohen et al. (2017) traced over-densities in the outer Galactic halo via the identification of variable RR-lyrae stars, whose heliocentric distance was directly obtained from their light curve.

- While inspecting HST photometry of the cluster NGC 6752 for the presence of white dwarfs, Bedin et al. (2019) discovered a dwarf spheroidal galaxy near the galaxy NGC 6744.

**5.3** *The bones of the Milky Way*

Determining the 3D structure of the Milky Way from our position very close to its mid-plane has been a perpetual challenge for astronomers. *Data exploration can provide a way around this*.

A good example is the discovery that the previously identified filamentary infrared dark cloud (IRDC) known as ***Nessie*** is in fact a very long structure that runs along the very center of the Scutum-Centaurus spiral arm of our galaxy, and acts as a "spine" that supports it. This realization came from re-analyzing archival *Spitzer* imaging data and combining it with kinematic properties derived from star-forming gas traced by CO and NH3 (Goodman et al. 2014).

The unexpected discovery of these dense structures that run along the spiral arms of the Milky Way and the realization that the Sun is offset by about 25 pc from the plane that contains them, provide a new advantageous perspective for the observer, and opens a new avenue for investigations of the 3D structure of our Galaxy.

**5.4** *Multiple stellar populations in globular clusters*

Globular Clusters were traditionally thought as single, first generation stellar populations. Observational evidence over the last decade has completely changed this view. For a given globular cluster, distinct populations enriched in He, N and Na, and distinct populations depleted in O and C, create complex patterns in the color-magnitude diagram that cannot be explained by individual stellar mixing and stellar evolution (Bastian & Lardo, 2018).

This discovery was highly unexpected. It came about through the exploration of large photometric datasets from HST observations, originally intended to characterize stellar evolution of a single populations.

Although several ideas have been put forward in order to explain this 'multiple populations problem', e.g., several bursts of star formation within the cluster, none of these theories can explain all the available observational evidence. Exploration of existing and upcoming pan chromatic surveys (Pan-STARRS, *Gaia*, LSST) are likely to boost a new way of discoveries and possibly provide some needed answers. Research in this particular area may provide new astrophysical insight about stellar evolution and star formation in the early universe.

**5.5** *Future data mining of the Hubble archive*

The great heritage of the *Hubble* Space Telescope includes multi-color imaging for hundreds of star clusters in the Milky Way and Magellanic Clouds, collected by different observing programs. The results from the many studies based on these images are hard or even impossible to compare, because they were processed using a heterogeneous set of tools to perform photometry and then interpreted using a variety of stellar population models (e.g. isochrones).

The *Hubble Legacy Archive* has produced a uniform set of data products from these data, so that now the photometry can be extracted homogeneously across the entire collection and then analyzed using one or several particular sets of stellar population models. Exploration of this uniform large archival data set may provide answers to questions such as: Why did the Large Magellanic Cloud not form any star clusters between 9 and 3Gyr ago? What is the binary star fraction in clusters, and is it affected by the clustered star formation? How did star clusters with multiple stellar population form?

# 6. Exploration in Galaxy Evolution

Given the nature of exploration, it is not possible to give definite questions that need to be addressed in the near future. Rather, we provide a few recent examples of (1) serendipitous unexpected discoveries (*unknown unknowns*) and their potential for changing established paradigms; and (2) new research avenues posed by asking very general questions (*known unknowns*), such as exploring the evolution of galaxies, which was the main driver of the COSMOS survey. We do not mean to provide an exhaustive survey of such discoveries, but only to illustrate our case with a few representative studies.

**6.1** *Unanticipated discoveries from improved observational capabilities*

Below we give four examples of **unanticipated** important discoveries, stemming from the opening of new observational windows due to the availability of new capabilities. The first three are well-known discoveries, which have led to the accepted scenario of joint galaxy-supermassive black hole evolution. The fourth example is a recent serendipitous discovery that may change the way we think about AGNs and their interaction with the cold ISM.

*Black Holes* became an observational reality with mass measurements of accreting BH X-ray binaries (first, Cyg X-1, Bolton 1972). The discovery of quasars and AGNs led to the successful model of accretion onto massive BHs in galaxy nuclei (Pringle et al 1973). With HST dynamical measurements, the widespread existence of supermassive BH became a generally accepted fact (Magorrian et al 1998), leading to our present understanding of galaxy-BH co-evolution.

*The hot gas of clusters and galaxies* is now a key 'observational' property of simulations of galaxy formation and evolution. It traces the universe's dark matter concentrations and is responsive to stellar and AGN feedback. However, until ~50 yrs ago nobody had thought of its existence. Observations with the first X-ray astronomy satellite, *Uhuru* (Giacconi et al. 1971) led to the discovery of the hot intra-cluster medium (Kellogg et al. 1971; Gursky et al. 1971). An imaging X-ray telescope, the *Einstein Observatory* (Giacconi et al 1979), was needed to discover hot halos in elliptical galaxies (previously believed to be devoid of ISM), and hot outflows from active star-forming galaxies and mergers (see review, Fabbiano 1989). With the spectral imaging capability of *Chandra* + ACIS, the interaction of AGNs with these hot halos (radio feedback; e.g. Paggi et al 2014) and the interaction of these hot halos with the intra-cluster medium are being mapped, providing a detailed picture of the dynamical universe.

*Extremely intense star-formation is a key stage of galaxy evolution.* This field of studies followed the discovery of extreme objects (ultraluminous infrared galaxies, ULIRGS) with the infrared mission IRAS (Soifer et al 1984). Their preponderance at high redshift was demonstrated by deep HST observations (Madau et al 1998). ALMA has uncovered a population of high-z, high-mass, dusty star-forming galaxies, with star-formation rates > 1000 Msol/yr (Blain et al 2002). These results are the foundation of our present understanding of galaxy merging evolution.

*The extended hard continuum (> 3 keV) and Fe Kα emission of Compton Thick AGNs.* This is a very recent, surprising discovery of X-ray sub-arcsecond spectral imaging, only possible with *Chandra* (Fabbiano et al 2017, Fig. 2). These extended components question the accepted model of simply torus-shrouded active nuclei and open a new avenue for exploring the observational intricacies of the AGN-galaxy interaction.

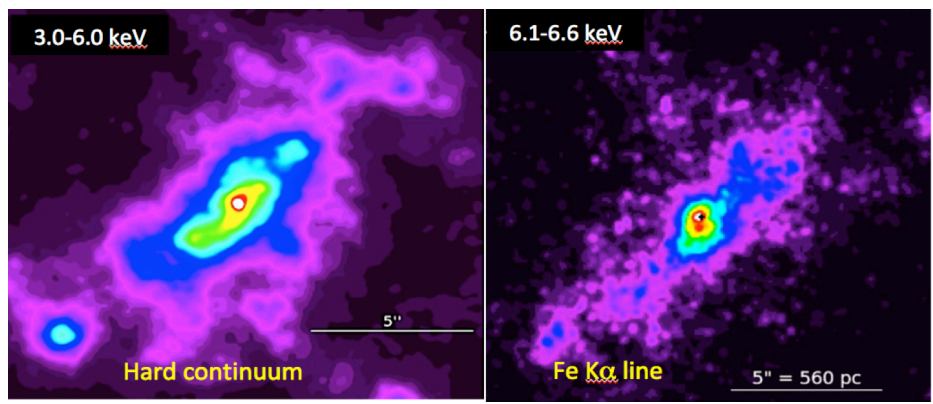

Figure 2. – ESO 428-G014 - > 2 kpc-scale hard continuum and ~1 kpc Fe K line emission (Fabbiano et al 2017)

**6.2** *New understanding from archival studies*

Several astronomical databases and data sets are freely available from astronomy archives and their community use is vigorous. Frequently these data sets are mined in combination with other databases or complemented with new observations. For example:

- The discovery of a new class of superluminous spiral galaxies, as optically luminous as first-ranked ellipticals in galaxy clusters, was made by mining multiwavelength data synthesized within the NASA/IPAC Extragalactic Database (NED) (Ogle et al. 2016).
- Workflows based on IVOA[2] standards and tools led Chilingarian et al. (2009; 2015) to expand to over 200 the sample of compact elliptical (cE) galaxies, of which only 6 examples were known, using archival HST images of nearby galaxy clusters and additional information from archives and databases, and mining SDSS and GALEX survey data. These works suggest formation by tidal stripping and cD ejection from host clusters and groups by three-body encounters.

A few illustrative recent examples based on the COSMOS multi-wavelength survey have probed the evolution of galaxies and nuclear activity out to large redshift:

- A connection between AGN activity and merging out to z~3.5: the fraction of Compton Thick (CT, $N_H > 10^{24}$ erg s$^{-1}$) AGNs in mergers/interacting systems increases with luminosity and redshift (Lanzuisi et al. 2018).
- Mature quenched bulges, discovered in star-forming galaxies at z~2, by mapping COSMOS galaxies with HST and VLT/SINFONI (Tacchella et al. 2015).
- A massive, dusty starburst in a galaxy protocluster at z = 5.7, serendipitously discovered in the COSMOS Field (complemented by ALMA and VLA), forming stars at a rate of at least 1500 M yr$^{-1}$ in a ~3 kpc compact region (Pavesi et al. 2018).
- Massive proto-clusters of galaxies at z~5.7 and z~4.6, discovered in the COSMOS field, using spectroscopic observations taken from Keck and the Visible Multi-Object Spectrograph (VIMOS) Ultra-Deep Survey (Capak et al. 2011; Lemaux et al. 2018).
- Over-massive BH (~$10^9$Msol) discovered in a $10^{10}$Msol star forming galaxy at z~3.3 (COSMOS + Keck), providing an example where BH growth may not be symbiotic with galaxy growth (Trakhtenbrot et al. 2015).
- The nature and luminosity function of galaxies with z~7-9 were explored using the COSMOS/UltraVISTA database complemented with HST imaging and Spitzer (Stefanon et al. 2017; Bowler et al. 2017).

---

[2] International Virtual Observatory Alliance; the forum for the development of the interoperability standards used by major astronomy datacenters (http://www.ivoa.net)

# 7. Exploration in Cosmology

Given the nature of exploration, it is not possible to give definite questions that need to be addressed in the near future. Rather, we provide a few recent examples of (1) serendipitous unexpected discoveries (*unknown unknowns*) and their potential for changing established paradigms; and (2) new research avenues posed by asking very general questions (*known unknowns*), such as constraining the cosmological parameters. We do not mean to provide an exhaustive survey of such discoveries, but only to illustrate our case with a few representative studies.

**7.1** *Unanticipated discoveries*

Cosmology is perhaps the best example of a field that was stimulated by observational discoveries, interplaying with theoretical models, beginning with the Hubble diagram and the Hubble-Lemaitre law. The current standard Lambda CDM (Cold Dark Matter) model of the observed Universe comes from the discovery of **dark matter** having several times the mass density of normal baryonic matter. The subsequent discovery of **dark energy** relegates baryonic matter to just below 5 % of the content of the Universe.

*Dark Matter* - The discovery of Dark Matter is based on serendipitous results of observational studies. The first – controversial - direct evidence was given by Fritz Zwicky in 1933 on the basis of the virial mass estimate in the Coma cluster while trying to refine the Hubble-Lemaitre relation. The idea did not get a broad acceptance at the time: many astronomers considered an alternative explanation to the mass problem by assuming that galaxy clusters are transient, non-virialized complexes. Convincing arguments for the need of the dark matter came from the studies of the rotation curves in spiral galaxies, but this study was originally undertaken to understand the dynamical properties and masses of galaxies throughout the Hubble sequence (Rubin et al. 1978). The gravitational attraction seen is clearly stronger than could be accounted for by the visible baryonic matter. There are expectations that this non-baryonic matter is in the form of some weakly interacting particle, but no direct detection of such particles has been achieved, and the arguments for detection of gamma-rays from their decay from astronomical sources have been questioned.

*Dark energy* was introduced by Albert Einstein in the form of a cosmological constant but estimates of its role based on the local and high redshift observations only started to appear in 1980s (Peebles 1988), to explain observational results: the growth rate of the large-scale structure seemed to require faster expansion of the Universe than predicted by the gravitational forces. *The convincing argument came from a totally independent line of study* – from Supernovae Ia (Riess et al. 1998, Perlmutter et al. 1999). SNIa are standardizable, and thus relatively easy to use for cosmology.

**7.2** *New Understanding from survey / archival studies: New cosmology probes*

In the last decade the values of the cosmological parameters in the standard Lambda CDM model have been determined precisely by combining the independent measurements of the Cosmic Microwave Background (CMB), SNIa, and Baryon Acoustic Oscillations (BAO). Combining the methods was essential, since there is considerable parameter degeneracy in any single model. This degeneracy is due to the limited range of redshifts in each method, the relatively large errors in individual measurements, and the limited number of measurements. Recently, tension is appearing between the newest measurements and the standard model. *These studies typically use and combine new extensive data bases, often obtained for different purposes.* Examples include:

- Riess et al. (2018), combining HST photometry and Gaia DR2 parallaxes of Cepheids to simultaneously constrain the cosmic distance scale and to measure the Gaia DR2 parallax zero point offset appropriate for Cepheids. They report a tension between the newest local measurements of the Hubble constant and those based on the CMB at close to the 4 sigma level.

- Risaliti & Lusso (2019), making use of a sample of 1,600 quasars from public archives (*XMM-Newton* and SDSS), complemented with new *XMM-Newton* observations, determined a deviation from the Lambda CDM model at the ~4 sigma level using the observed non-linear relation between AGN broad band UV/X-ray flux ratio and UV luminosity.

- Emami et al. (2019), studying the clustering amplitude of a complete sample of 7143 clusters in the Sloan survey, report that the observed correlation length exceeds pure CDM simulation prediction by ~6% for the standard Plank-based parameters. This excess may be explained by free streaming of light neutrinos.

These discrepancies suggest the possibility of new physics beyond the standard ΛCDM cosmology. The combination of different, sometimes new, approaches, supported by an increased data quality and sample statistics, is the way forward to solve the dark matter and dark energy problems. In the future, we expect cosmological results from a range of astronomical measurements, making use of publicly released surveys and using tailored analysis workflows, including weak lensing (e.g., Mandelbaum 2018), strong gravitational lensing events (e.g., Cao et al. 2015), Supernovae II (de Jaeger 2015), Active Galactic Nuclei reverberation mapping (AGN; Cackett et al. 2007, Watson et al. 2011, Haas et al. 2011), galaxy clusters through combined X-ray and Sunyaev-Zeldovich effect (e.g., Bonamente et al. 2006), gamma-ray bursts (Schaefer 2007), and gravitational wave events (Abbott et al. 2017). While the methods to use these probes are still in development, some promising results are coming, paving the way for further advances in the 2020s

# 8. Increasing the Discovery Space

**8.1** *Observing facilities that expand boundaries*

Any new observing facilities/missions for the next decade should significantly improve performance in some key metric (e.g., energy range, sensitivity, exposure time, angular resolution, higher dimensional data, rapid response), and be well characterized and calibrated, so to provide flexibility for new observing avenues. *Hubble*, *Spitzer* and *Chandra* provide examples in the discovery of Dark Energy, the detection of z=11 galaxies, and the nature of Dark Matter (Bullet Cluster), respectively. Beyond hardware capabilities these discoveries require: mission longevity, community driven science, high-quality data products in readily accessible, interoperable, archives and a well-supported user/observer community.

**8.2** *Multi-wavelength and multi-messenger capabilities*

Many historical examples also demonstrate a strong synergy between different wavebands and messengers. Having contemporaneous access to the entire electromagnetic spectrum was vital to finding the first counterpart to a gravitational wave source, for example. This multi-wavelength coverage of the sky that we are currently enjoying needs to be preserved.

**8.3** *Curated Data Archives and Powerful Data Analysis tools*

These new facilities will generate increasingly larger and complex multi-wavelength and multi-messenger data sets and catalogs. These data will need to be properly reduced and curated to fully enable their discovery potential. *Archives must provide both easy access to these data and (with the community) the means to exploit them*.

The above goals translate into:

(1) Ensure that any operational (old and new) facility/mission explicitly include in their scope the proper processing of software so to produce well documented and calibrated data products, as well as the capability for data recalibration and reprocessing.

(2) Organize these data products in well-maintained archives, following the International Virtual Observatory Alliance (IVOA)[3] standards, so to allow a basic level of access and *interoperability*, as well as *repurposing.* Much of this is already in place in the NASA archives, and they are collaborating in extending and evolving these capabilities to meet the demands of new data types and research methods through the 2020s. Data products should be replicable and reproducible, ranging from basic observation data to high-level aggregated data and catalogs.

---

[3] The forum for the development of the interoperability standards used by major astronomy datacenters (http://www.ivoa.net)

(3) Ensure that data centers engage in the development and refinement of interoperability standards, via the well-established processes of the IVOA, and work with groups such as *Astropy*[4] to ensure support for these standards in present in community developed, open source software.

(4) Ensure that new facilities are adequately supported. *New facilities* (Sections 3.1, 3.2) *will demand a transformation in the way data are analyzed*. The early phases of this transformation are already underway (e.g., the use of *Python* as an environment, cloud computing). But, resources must be made available for full development, which will demand remote Science Platforms[5] and Server-side analytics[6], implementation of complex fault-tolerant workflows, data mining and machine learning, and advanced visualization.

(5) Foster the development of *next generation* interoperable, user-friendly visual interfaces, data mining tools, the ability to construct and implement analysis workflows easily, both via visualization and scripting, and the ability to work with data both locally and remotely (current-generation well-know examples include TOPCAT, DS9 and CSCView).

(6) Support interdisciplinary research in astrostatistics and astroinformatics and the transfer of methods from the statistics, computer science, and machine learning communities, for development and application of innovative data analysis methods and algorithms.

(7) Ensure that facilities and archives participate in curation efforts and initiatives to link together datasets, related ancillary data (e.g., atomic and molecular databases), objects, and the literature.

Data are an important legacy of major astronomical facilities, and proper data maintenance will insure that new science will be produced for the future, even after the first crop of scientific papers and discoveries have been published. Statistics of data usage from the NASA archives demonstrate that archival data is used for new published scientific work several times (Fig. 3).

---

[4] http://www.astropy.org/acknowledging.html
[5] See LSST Science Platform Design document https://ldm-542.lsst.io
[6] NASA Big Data Task Force (https://science.nasa.gov/science-committee/subcommittees/big-data-task-force)

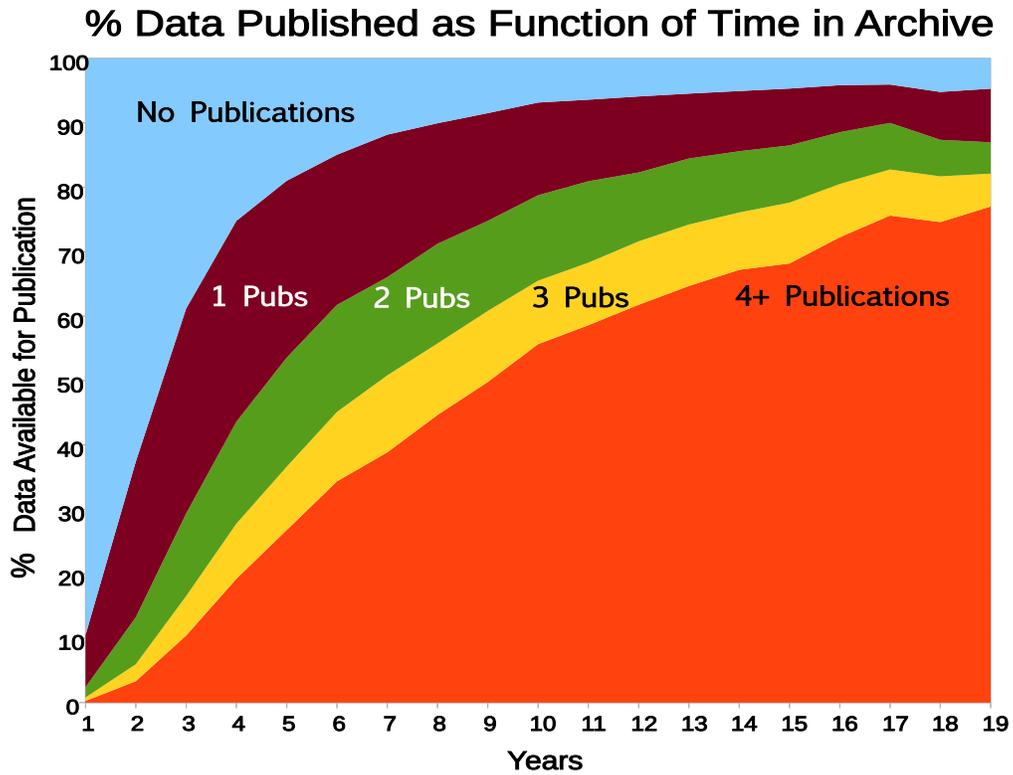

Figure 3. – Percentage of *Chandra* exposure time published versus years in the archive. The scientific use of archival *Chandra* data is increasing with time in the public archive. For example, 19 years from launch, ~75% of the observation have been published in more than 4 papers. A similar trend is observed for the HST data.

## 9. Conclusions

We propose *exploration* as the central question for the Decadal Committee's discussions. The history of astronomy shows that paradigm-changing discoveries were not driven by well-formulated scientific questions, based on the knowledge of the time. They were instead the result of the increase in discovery space fostered by new telescopes and instruments. An additional tool for increasing the discovery space is provided by the analysis and mining of the increasingly larger amount of archival data available to astronomers. We urge the Decadal Committee to **(1)** *keep multi-wavelength and multi-messenger exploration center stage* in their deliberations of new facilities, including consideration for flexible and well-calibrated modes of operation that could foster adaptation for use with new discovery space; and **(2)** *recognize the importance of data and their stewardship, and computational services*, as major elements of any new scientific development for the next decade. **Revolutionary observing facilities, and the state-of-the-art astronomy archives needed to support these facilities, will open up the universe to new discovery.**